\documentclass[sn-apa]{sn-jnl}
\usepackage{comment}
\usepackage{xcolor}
\usepackage{array}
\usepackage{multirow}

\jyear{2023}%

\theoremstyle{thmstyleone}%
\theoremstyle{thmstyletwo}%

\theoremstyle{thmstylethree}%

\begin{document}

\title[Detecting Anomalous Cryptocurrency Transactions]{
Detecting Anomalous Cryptocurrency Transactions: an AML/CFT Application of Machine Learning-based Forensics
}


\author*[1,3]{
    \fnm{Nadia} \sur{Pocher}
}
\email{nadia.pocher@uab.cat}

\author[2,3]{
    \fnm{Mirko} \sur{Zichichi}
}
\email{mirko.zichichi@upm.es}

\author[3]{
    \fnm{Fabio} \sur{Merizzi}
}
\email{fabio.merizzi@studio.unibo.it}

\author[3]{
    \fnm{Muhammad Zohaib} \sur{Shafiq}
}
\email{muhammad.shafiq6@studio.unibo.it}

\author[4]{
    \fnm{Stefano} \sur{Ferretti}
}
\email{stefano.ferretti@uniurb.it}

\affil[1]{
    \orgdiv{Institute of Law and Technology, Faculty of Law}, 
    \orgname{Universitat Autònoma de Barcelona},
    \orgaddress{
        \city{Bellaterra}, 
        \postcode{08193}, 
        \country{Spain}
   }
}

\affil[2]{
    \orgdiv{Ontology Engineering Group}, 
    \orgname{Universidad Politécnica de Madrid},
    \orgaddress{
        \street{ETSIINF}, 
        \city{Boadilla del Monte (MD)}, 
        \postcode{28660}, 
        \country{Spain}
    }
}

\affil[3]{
    \orgdiv{Department of Computer Science and Engineering and Department of Legal Studies}, 
    \orgname{University of Bologna},
    \orgaddress{
        \city{Bologna}, 
        \postcode{40126}, 
        \country{Italy}
    }
}

\affil[4]{
    \orgdiv{Dipartimento di Scienze Pure e Applicate}, 
    \orgname{University of Urbino ``Carlo Bo''},
    \orgaddress{
        \street{Piazza della Repubblica, 13}, 
        \city{Urbino}, 
        \postcode{61029}, 
        \country{Italy}
    }
}

\abstract{
    In shaping the Internet of Money, the application of blockchain and distributed ledger technologies (DLTs) to the financial sector triggered regulatory concerns. Notably,
    while the user anonymity enabled in this field may safeguard privacy and data protection, the lack of identifiability hinders accountability
    and challenges the fight against
    money laundering and the financing of terrorism and proliferation (AML/CFT).
    As law enforcement agencies and the private sector apply forensics to track { crypto transfers across ecosystems that are socio-technical in nature}, this paper focuses on the growing relevance of these techniques { in a domain where their deployment impacts the traits and evolution of the sphere}.
    In particular, this work offers contextualized insights into the application of methods of machine learning and 
    transaction graph analysis.
    Namely, it analyzes a real-world dataset of Bitcoin transactions represented as a directed graph network through various 
    techniques. The modeling of blockchain transactions as a complex network suggests that the use of
    graph-based
    data analysis methods 
    can help classify transactions and identify illicit ones.
    Indeed, this work shows that the neural network types known as Graph Convolutional Networks (GCN) and Graph Attention Networks (GAT) are a promising AML/CFT solution. 
    Notably, in this 
    scenario GCN outperform other classic 
    approaches { and GAT are applied for the first time to detect anomalies in Bitcoin. Ultimately, the paper upholds the value of public-private synergies to devise forensic strategies conscious of the spirit of explainability and data openness.}
}


\maketitle

\section{Introduction}\label{sec:intro}

Over the last 15 years, the application of blockchain and distributed ledger technologies (DLTs) to the financial domain has generated an enthusiastic hype~\citep{ali2020state}. Building on years of research in distributed systems and cryptography, the launch of Bitcoin~\citep{Nakamoto2008} showed it is possible to reliably record information ({\em e.g.}, transactions) 
without trusting a centralized
party. This opened the way to peer-to-peer transfers and direct participation 
in a digital global economy. 
However, the features of disintermediation and perceived anonymity of this Internet of Money~\citep{Antonopoulos2017a} 
cause regulatory unease.\footnote{The Internet of Money is neither a legal nor a technical definition; in this work, the term refers to the entire set of cryptocurrency ecosystems, thus including the part of the Internet of Value~(\cite{Tapscott2019}) that relates to payment tokens.}
Indeed, they defy accountability and fuel fears of exploitation for 
illicit purposes~\citep{chang2020blockchain}. 
As confirmed by industry estimates, 
{ in 
2022 the volume of crypto-related illicit activity
hit USD 20.6 billion and increasingly involves decentralized finance (DeFi)~\citep{ChainalysisTeam2023}.}
This challenges the fight against
money laundering and the financing of terrorism and proliferation (AML/CFT). 

The AML/CFT framework consists of a set of laws, regulations and procedures that aim to protect the integrity of the financial system mainly by making the concealment of the origin of illicit profits significantly troublesome 
~\citep{doubleblind}. 
Since the identification of customers and counterparties is a key part of
AML/CFT compliance 
for entities such as financial institutions and cryptoasset service providers, some features of the Internet of Money that hinder identifiability emerge as problematic.
However, 
crypto-related laundering 
appears heavily concentrated: most value originating from illicit addresses is seemingly sent to few services, often built for criminal purposes~\citep{ChainalysisTeam2023}.
This suggests the key role of effective, and possibly efficient, classification of transactions performed by/received from specific entities to detect and investigate illicit activities in the sphere at hand.


In this context, 
the picture of 
untraceable cryptocurrency transfers and individual freedom from governmental control warrants a two-fold interpretation: 
while user anonymity can safeguard privacy and data protection,
lack of identifiability hampers investigation, enforcement, and accountability. 
Two sets of mutually influencing socio-technical events emerged: 
law enforcement agencies and private sector providers of RegTech
solutions started exploring techniques to ``follow the money'' across blockchain ecosystems~\citep{Meiklejohn2016, Biryukov2019, Chen2019, Bartoletti2020, Moreno2016, Lischke2016}, while the unveiled insufficiency in Bitcoin's anonymity spurred altcoin projects ({\em e.g.}, Monero, ZCash) 
to implement advanced cryptographic methods that require new analytical tools.\footnote{The term ``RegTech'', short for ``regulatory technology'', refers to the use of new technologies to aid regulatory and compliance processes, mainly through FinTech software applications.}

Against this backdrop, in this paper we focus on the value of intelligence techniques to provide insights into the Internet of Money's ecosystems, with specific regard to machine learning techniques, network and transaction graph analysis~\citep{Weber2019, Fleder2015, Ober2013, Wu2021}.
We first provide a background on a notion of anonymity that is specific to the Internet of Money and on the interplay of AML/CFT and blockchain forensics. Consequently, we focus on the anomaly detection approaches that led to our experiments. 
{
In particular, we employed a dataset obtained from a set of Bitcoin transactions, represented as a directed graph network~\citep{Weber2019}. 
}
The modeling of Bitcoin transactions as a complex network fosters the use of specific graph-related analysis techniques, which usually help identify peculiar nodes of a network 
(\cite{doubleblind}).
As per our central hypothesis, since money laundering involves transaction
flow relationships between entities creating a graph structure, AML/CFT analytics could benefit from novel graph analysis techniques in machine learning, namely Graph Convolutional Networks (GCN) and Graph Attention Networks (GAT).

The results of our experiments show that GCN and GAT neural network typologies are promising solutions for AML/CFT. This is in opposition to a state-of-the-art work in which a baseline supervised learning algorithm, {\em i.e.}, non-graph-based, such as the Random Forest, provided the best performances~\citep{Weber2019}.
Thus, we underline the value of experimenting with techniques based on machine learning and transaction graph analysis and their 
combinations. 
We contextualize our argument vis-à-vis the amount and complexity of crypto transaction data 
and the specifics of AML/CFT anomaly indicators. 
We do so by considering the need to mitigate the shortcomings of rule-based regimes, explainability aspects, and the urgency to engage in research informed by an interdisciplinary methodology.

To summarize, the main contribution of this work is twofold:
\begin{itemize}
    \item We show how modeling blockchain transactions as complex networks is conducive to the subsequent application of specific graph-based learning approaches for anomaly detection purposes. In particular, our experiments show how the GCN model generates better results than other machine learning methods. Notably, it seems to outperform  state-of-the-art implementations in classifying illicit transactions;
    \item Our work heeds a compound of technical, operational and regulatory viewpoints when considering the benefits of machine learning for AML/CFT anomaly detection. This allows us to account for the need for interpretability and explainability, as well as the effectiveness and efficiency of the deployed approaches.\footnote{Research into the regulatory impacts of the explainability and interpretability of AI applications is vast and detailed; in light of the scope of this work, we perform inevitable simplifications.} This methodology displays the value of cross-disciplinary models to improve accuracy, significantly aiding compliance and investigation, reducing false positives and over-reporting. 
\end{itemize}

The remainder is structured as follows. Section II provides a conceptual background on Bitcoin's pseudonymity and insights into the relationship between AML/CFT and forensics. Section III explores related work. Section IV takes a context-specific approach to outline anomaly detection techniques. In Sections V and VI, 
we present and discuss our study on 
machine learning-based AML/CFT classification methods. Section VI concludes the paper. 

\section{Background}\label{sec:back}

{
In this section, we discuss  key aspects related to pseudonymity and de-anoymization, followed by a discussion on AML/CFT and blockchain forensics. 

Preliminarily, we point out that in this work the terms DLT and blockchain are used as synonyms. As is known, while the term DLT refers to a generic idea of distributed ledger, regardless of its implementation, a blockchain is a specific form of DLT in which transactions are stored as a sequence of blocks. Even if the term blockchain is more specific, 
it is more popular and often used in a wider sense. 
In this work, the way the DLT stores transactions in the ledger does not influence our model. Indeed, our approach considers the graph generated by transactions, {\em i.e.}, a direct link from a transaction, say $t_1$, to another $t_2$, exists if the money earned in $t_1$ is spent in $t_2$. Thus, our study focuses on a layer that is higher than the ledger where transactions are stored.
}

\subsection{Pseudonymity and De-Anonymization}\label{sec:pseudo} 

Untraceability of payments was among the goals of the Internet of Money~\citep{DeFilippi2018}. However, in concrete terms the latter is populated by many socio-technical notions of anonymity and transparency
~\citep{doubleblind}. 
{ Following an holistic interpretation, cryptocurrencies are generated and exchanged within socio-technical systems that, as such, comrpise of interdependent technology and human systems~\citep{Baxter2011, Desmond2019}. Hence, their characteristics are influenced by social and technical aspects.}
Since a literature review~\citep{Amarasinghe2019} falls outside the scope of our work, we heed the specific understanding that anonymity in the Internet of Money ``means being able to conduct a financial transaction without anyone besides the sender and the receiver being able to identify the parties involved''~\citep{Edmunds2020}. 
Indeed, it is a common blockchain goal to combine user anonymity and transparency of operations ({\em i.e.}, ledger transparency), and a public blockchain is structurally designed to enable anonymous peer-to-peer transfers~\citep{Quiniou2019}. 

There is wide agreement that Bitcoin is pseudonymous, and not anonymous~\citep{Biryukov2019, Li2019, Berg2019}.
Pseudonymity refers to the use of pseudonyms as identifiers, and a pseudonym is a subject's identifier other than the subject's real name~\citep{Pfitzmann2010}. 
In most blockchain systems, public-private key pairs 
uniquely identify wallet holders~\citep{Wang2020}. Hence, in a crypto transaction, addresses ({\em i.e.}, public keys) perform the function of  usernames. It follows that senders and recipients are pseudonymous, not anonymous, when their address identifies them. 
However, this is not sufficient from a regulatory perspective because pseudonyms alone do not ensure accountability. Indeed, when AML/CFT rules require identification, they refer to real-world identities.

In principle, a currency scheme aims to prevent that the transaction history of its units can be retraced. If it is possible to associate a coin with its past exchanges, the currency's fungibility is threatened, and its nominal value is affected.
Because Bitcoin's features seemed insufficient, new techniques have been embedded into anonymity-enhanced currencies (AECs), also known as ``privacy coins'', to bypass regulatory constraints and surveillance. 
They deploy privacy-enhancing technologies such as zero-knowledge proofs (ZKPs).
{ Concurrently, even if the most common way to buy and exchange cryptocurrencies still relies on centralized exchanges, the Internet of Money is witnessing the emergence of DeFi applications,\footnote{{ DeFi was defined as an ``ecosystem of financial services realised through smart contracts deployed on public distributed
ledgers''~\citep{Amler2023}, where the role of intermediaries is replaced by self-executing
computer code~\citep{Katona2021}. Nonetheless, the levels of decentralization of the relevant projects vary and are debated \citep{Barbereau2022}.}} such as stablecoin projects ({\em e.g.}, DAI), lending platforms ({\em e.g.}, Aave, Compound), decentralized exchanges ({\em e.g.}, Uniswap, Pancakeswap)~\citep{Amler2023, Aramonte2021, Katona2021}. The total
value of DeFi projects reportedly amounted to USD 1 billion in January 2020, USD 27 billion
in January 2021, USD 60 billion in April 2021, and USD 40 billion in November 2022~\citep{ChainalysisTeam2022}.
} 

Meanwhile, the private sector and law enforcement professionals have devised strategies to trace transfers in the Internet of Money. The end goal of these intelligence methods is to match users, definitively or statistically, to transactions performed by crypto-addresses -- {\em i.e.}, to connect pseudonyms to real-world identities -- leveraging unique identifiers. These techniques were originally labeled ``blockchain forensics'', as they were informed by the specificities of blockchains and defined as the use of science and technology for the sake of investigation and fact-establishment in a court of law, primarily dealing with recovering and analyzing the evidence on blockchain ledgers~\citep{Phan2021}. Later, analytic solutions started to be requested by regulated entities. Although they have been mostly tested on the Bitcoin network, data-exploitation strategies have been deployed on Ethereum~\citep{Chen2019, Bartoletti2020, Li2021, Moreno2016}, and on non-blockchain DLTs~\citep{Tennant2017, Ince2018}.
{ Evidently, both obfuscation and traceability are not only endeavours pursued by actors belonging to the crypto sphere, but also activities that influence the overall anonymous or transparent socio-technical character of the domain.}
    
Since identifiers ({\em i.e.}, addresses/public keys) can be leveraged to connect transactions to their history, Bitcoin's pseudonymity generates an inherent tension between anonymity and accountability~\citep{Yin2019}. However, unless they are associated with additional data, identifiers do not reveal personally identifying information~\citep{Wang2020}. Hence, pseudonymity does not imply identifiability, which is subjective: a pseudonymous subject is identifiable only if a specific actor can discover its real-world identity.
This is crucial because, in the Internet of Money, there are both (a) actors, such as authorities and cryptoasset service providers, that seek to achieve identification, and (b) strategies employed at various levels to avert it, {\em e.g.}, advanced cryptography and virtual private networks.
This is an example of how technology can both foster new pathways to accountability and disrupt data retrievability.
In this context, the transparent nature of (public) blockchains makes them vulnerable to insufficient data privacy, de-anonymization attacks, and surveillance. While de-anonymization is often perceived negatively, it can be applied for investigative purposes and to comply with rules that aim to mitigate specific risks, such as AML/CFT.

\subsection{AML/CFT and Blockchain Forensics} \label{sec:aml}

The first concern of cryptocurrency misuse originated from transactions on the dark web. While a range of technologies aids darknet operations, cryptocurrencies, mainly Bitcoin and Monero, play a crucial role by facilitating payments~\citep{Akhgar2021}.
While Bitcoin is still the major player, used by 93\% of darknet markets, the adoption of Monero is increasing: 67\% of platforms supported it in 2021 vis-à-vis 45\% in 2020, and some support it on an exclusive basis~\citep{ChainalysisTeam2022}.
Nonetheless, the public perception of crypto-related laundering is likely inflated. 
{
Indeed, even if the value of illicit crypto transactions reached an all-time high in 2022, hitting USD 20.6 billion, it accounts only for 0.24\% of crypto activity~\citep{ChainalysisTeam2023}, and remains small when compared with criminal activities involving fiat currencies~\citep{Goforth2020, CipherTrace2021}.\footnote{{ It is worth noting that in 2022 the share of crypto activity associated with illicit activity rose for the first time since 2019. However, 43\% of the illicit transaction volume is linked to sanctioned entities. Notably, for the most part, to the crypto exchange Garantex~\citep{ChainalysisTeam2023}}}}


Since the risk-based approach informs AML/CFT obligations, regulated entities must tune compliance efforts: stricter measures if risk factors are higher. 
The end goal is to draw authorities' attention when suspicions of illicit activities arise by filing a
report when the entity knows, suspects, or has reasonable ground to suspect the given funds are the proceeds of a criminal activity or are related to terrorist financing~\citep{FATF2022, EuropeanParliament2018}.
Generally, AML/CFT duties apply to crypto-transactions,
and cryptoasset service providers
are increasingly regulated. In the EU, the 5\textsuperscript{th} AML Directive~\citep{EuropeanParliament2018} first targeted these activities, and the regime is evolving with the 
AML Package~\citep{EuropeanCommission2021package}. 

Even if
{ the ledger transparency featured by public blockchains}
mitigates the risk of fraudulent behavior,
the technology is vulnerable to unpredictable exploitation methods~\citep{xu2016, sym14020328}. This prompted the development of specific techniques of anomaly detection.
In this field, the Internet of Money's opaque reputation appears paradoxical since it provides a huge amount of open-source intelligence 
-- {\em e.g.}, it is possible to extract data from a given transaction and retrieve the history of an address, while methods using networks created by transactions ({\em i.e.}, ``transaction flow analysis'') can 
define patterns to pinpoint suspected addresses~\citep{Wu2021}.
Different analytic techniques have been refined over time~\citep{Yin2019}, and mostly rely on statistical approaches -- {\em e.g.}, the re-use of an account for more transactions or the co-use of more accounts for a single transaction can lead to matching more accounts to the same user~\citep{Li2021}.

{ Starting from 2020, a surge of ransomware attacks highlighted regulatory shortcomings concerning the complex development of the Internet of Money. Indeed, as the latter becomes populated by AECs and other services that 
increase obfuscation, the risks of fraud increase. 
More recently, in 2022 hackers stole USD 3.1 billion from DeFi protocols, exploiting their transparency -- {\em i.e.}, typically, DeFi transactions happen on-chain and the smart contract code is publicly viewable. This amount accounts for 82\% of all crypto funds stolen by hackers. In the same year, crypto mixers processed USD 7.8 billion, 24\% of which originated from illicit addresses~\citep{ChainalysisTeam2023}. 

To guide regulated entities in the management of their exposures, several authorities publish red flag/risk indicators to guide compliance and supervision.} Notably, in the indicators published by
the global AML/CFT standard-setter Financial Action Task Force (FATF)
there is a section on anonymity risks~\citep{FATF2020},\footnote{The report targets six types of indicators, relating to (i) transactions, (ii) transaction patterns, (iii) anonymity, (iv) senders/recipients, (v) funding/wealth at source, (vi) geographic risks.} updated in 2021~\citep{FATF2021review}. 
%
%
Although a transaction's anonymity level is insufficient to suggest the transfer is suspicious, the FATF underlined inherent issues of privacy-enhancing technologies implemented by privacy coins, such as ZKPs~\citep{FATF2020}. At the same time, a range of institutions highlighted the risks caused by unhosted wallets~\citep{Europol2020, ChainalysisTeam2023}. 

Against this backdrop, forensic methods provide a wide range of information that emerges as pivotal for investigation, compliance and supervision.
Their value is displayed by the debate on the crypto travel rule, pursuant to which regulated entities must identify originators and recipients of crypto-transfers to guarantee traceability. 
In principle, this is just an expansion of data sharing measures previously applicable only to wire transfers, as required by the FATF Standards and by EU measures part of the AML Package. However, the reactions to the crypto travel rule exemplify the tension between the Internet of Money and an intermediary-based regulatory framework that still has to capture the specifics of peer-to-peer transfers and decentralized platforms. 
Accordingly, the industry denounces the absence of global standards and technical solutions to underpin effective and affordable compliance.

\section{Related Work}\label{sec:related}

While we do not aim to offer a review of the
techniques of cryptocurrency forensics,
in this section we describe a few works 
that provided an application to the concepts introduced above.
In blockchain analytics, various methods aim to link pools of addresses and transactions. 
They can deploy clustering techniques to group addresses owned by the same user~\citep{Neudecker2017, Ince2018, Wu2021} and also leverage transaction graphs to explore the features of the network~\citep{Weber2019, Fleder2015, Ober2013, AlJawaheri2020}. Some of these approaches aim to identify idioms of use in the network that can erode anonymity~\citep{Meiklejohn2016}, while others screen transactions to/from crypto-wallets to classify transactions as licit or illicit~\citep{Weber2019}.
In principle, these tools do not directly try to link addresses and transactions to real-world identities. However, if one of them is de-anonymized (in other ways), they allow to de-anonymize the whole cluster, as the cluster database allows fast correlation. Likewise, the goal usually is not to identify transaction patterns, but to allow that when an addresses is suspected other addresses of the same cluster can be suspected as well~\citep{Wu2021}.  

Clustering methodologies are based on heuristic models~\citep{Lischke2016, Reid2013},
such as:
if two/more addresses are inputs to the same transaction, they are controlled by the same user~\citep{Meiklejohn2016}.
In wallet-closure analysis the 
heuristics are applied to establish a unique mapping between addresses and an identity~\citep{AlJawaheri2020}. 
In behavior-based clustering~\citep{Yin2019},  addresses are grouped based on patterns such as transaction values~\citep{Amarasinghe2019}. A study performed by~\citep{Androulaki2013} showed this could unveil the profiles of 40\% of Bitcoin users despite privacy measures. 

On the application level, analytic techniques can exploit the possibility to correlate transactions with
users' information on social media. Frequently, users post their addresses ({\em e.g.}, to receive donations) but also reveal personal information ({\em e.g.}, contact information, age, location)~\citep{AlJawaheri2020}. In this respect, transaction fingerprinting methods can make use of off-network data~\citep{Reid2013}, which is also leveraged by web-scraping and Open Source Intelligence tools. The authors in~\citep{Fleder2015} annotated the transaction graph by linking user pseudonyms to online identities collected from social media and developed a graph-analysis framework to summarize and cluster users' activity to link identities and transactions. 

Specific methods target mixing services~\citep{Wu2020}, {\em i.e.}, the ones that shuffle coins by sending them to different addresses to obfuscate the flow.
Although third-party services act as centralization points, thus aiding traceability, new disintermediated methods such as 
CoinJoin~\citep{AlJawaheri2020} deploy more sophisticated shuffling approaches. In this context, an important role is played by peer-to-peer cross-chain transfers, and a relatively new subset of analytic efforts aims to trace cross-currency transfers through exchanges such as ShapeShift~\citep{AlJawaheri2020}. The authors in~\citep{HarriganMartin2016TUEo} clustered the addresses of the whole Bitcoin blockchain to show that the methodology remains effective despite mixed transactions.

Another line of forensic research, further discussed in section \ref{sec:anomaly},
is based on machine learning. 
The authors in~\citep{Yin2019} presented a supervised learning-based approach to de-anonymize the Bitcoin blockchain to predict the type of entities yet not identified. They built classifiers concerning 12 categories
and concluded that it is possible to predict the type of an entity. 
To do so, they collaborated with the analytic company Chainalysis that provided the data and had previously clustered, identified and categorized a considerable number of addresses manually or through clustering techniques.
They show two examples, one where they predict a set of 22 clusters suspected to be related to criminal activities, and another where they classify 153,293 clusters to provide an estimation of Bitcoin activity. 
Furthermore, they concluded it is possible to predict if a cluster belongs to predefined categories such as exchange, gambling, merchant services, mining pool, mixing, ransomware, and scam.

{\renewcommand\normalsize{\footnotesize}%
\normalsize
\begin{table*}[ht]    
    \caption{
Summary of the features and comparison of related works with our work.}
    \centering
    \begin{tabular}{ >{\centering\arraybackslash}m{1.2cm}  >{\centering\arraybackslash}m{2cm}  >{\centering\arraybackslash}m{3cm}  >{\centering\arraybackslash}m{3.9cm} }
     \hline
     \textbf{Work} & \textbf{Methodology} & \textbf{Algorithms} & \textbf{Results}\\ 
     \hline
     \hline
     
     \cite{Reid2013} & 
     \shortstack{\vspace{0.01cm}\\ Network Analysis} & 
     \shortstack{\vspace{0.01cm}\\ flow analysis +\\ off-network information} & 
     \shortstack{\vspace{0.01cm}\\ associate addresses with each\\ other and with external\\ identifying information} \\ 
     \hline
     
     \cite{Fleder2015} & 
     \shortstack{\vspace{0.01cm}\\ Network Analysis} & 
     \shortstack{\vspace{0.01cm}\\ flow analysis +\\ web scraping} & 
     \shortstack{\vspace{0.01cm}\\ link illicit activities to\\ online identities } \\ 
     \hline
     
     \cite{Wu2021} & 
     \shortstack{\vspace{0.01cm}\\ Network Analysis} & 
     \shortstack{\vspace{0.01cm}\\ safe Petri Net-based\\ cluster analysis} & 
     \shortstack{\vspace{0.01cm}\\ find suspected addresses} \\ 
     \hline
     
     \cite{AlJawaheri2020} & 
     \shortstack{\vspace{0.01cm}\\ Network Analysis} & 
     \shortstack{\vspace{0.01cm}\\ wallet-closure analysis} & 
     \shortstack{\vspace{0.01cm}\\ infer links between Bitcoin\\ users and hidden services} \\ 
     \hline
     
     \cite{HarriganMartin2016TUEo} & 
     \shortstack{\vspace{0.01cm}\\ Network Analysis} & 
     \shortstack{\vspace{0.01cm}\\ address-clustering analysis} & 
     \shortstack{\vspace{0.01cm}\\ identify super-clusters} \\ 
     \hline
     
     \cite{sun2021cubeflow} & 
     \shortstack{\vspace{0.01cm}\\ Graph Analysis} & 
     \shortstack{\vspace{0.01cm}\\ flow-based graphs analysis\\ with coupled tensors} & 
     \shortstack{\vspace{0.01cm}\\ anomalous transactions\\ detection\\ FAUC metric 0.94} \\ 
     \hline
     
     \cite{li2020flowscope} & 
     \shortstack{\vspace{0.01cm}\\ Graph Analysis} & 
     \shortstack{\vspace{0.01cm}\\ theoretical flow-based\\ multipartite graphs analysis} & 
     \shortstack{\vspace{0.01cm}\\ anomalous transactions\\ detection\\ FAUC metric 0.96} \\ 
     \hline
     
     \cite{Yin2019} & 
     \shortstack{\vspace{0.01cm}\\ Machine Learning} & 
     \shortstack{\vspace{0.01cm}\\ supervised learning-based\\ (baseline)} & 
     \shortstack{\vspace{0.01cm}\\ predict type of \\ yet-unidentified entity\\ F1score 0.796 (GradientBoosting)} \\ 
     \hline
     
     \cite{Weber2019} & 
     \shortstack{\vspace{0.01cm}\\ Machine Learning\\ + Graph Analysis} & 
     \shortstack{\vspace{0.01cm}\\ supervised learning-based\\ (baseline + GCN)} & 
     \shortstack{\vspace{0.01cm}\\ predict illicit transactions\\ F1score 0.796 (Random Forest)} \\ 
     \hline
     
     \cite{Eddin2021} & 
     \shortstack{\vspace{0.01cm}\\ Machine Learning\\ + Graph Analysis} & 
     \shortstack{\vspace{0.01cm}\\ supervised learning-based\\ (baseline + triage model)} & 
     \shortstack{\vspace{0.01cm}\\ reduce the number of false\\ positives by 80\%} \\ 
     \hline
     
     \cite{oliveira2021guiltywalker} & 
     \shortstack{\vspace{0.01cm}\\ Machine Learning\\ + Graph Analysis} & 
     \shortstack{\vspace{0.01cm}\\ supervised learning-based\\ (baseline + GuiltyWalker)} & 
     \shortstack{\vspace{0.01cm}\\ predict illicit transactions\\ F1score 0.85 (Random Forest)} \\ 
     \hline
     
     \textbf{Ours} & 
     \shortstack{\vspace{0.01cm}\\ Machine Learning\\ + Graph Analysis} & 
     \shortstack{\vspace{0.01cm}\\ supervised learning-based\\ (baseline + GCN + GAT)} & 
     \shortstack{\vspace{0.01cm}\\ predict illicit transactions\\ F1score 0.844 (GCN)} \\ 
     \hline
    \end{tabular}
    \label{tab:tableComparison}
\end{table*}
}

Machine learning solutions benefit from constructing multiple graph types from blockchain data, {\em e.g.}, a blockchain account (or a group of) is a node, and a single transaction between two accounts is an edge. An edge's weight is then defined as the aggregate transaction volume over a period of time. The latter is the predominant crypto-related forensic method seen in Section~\ref{sec:aml}~\citep{Weber2018}. Relatedly, the authors in~\citep{Weber2019} benchmarked GCN against various supervised methods. In contrast, the authors in~\citep{Eddin2021} extended their work to reduce false alerts through supervised learning methods in a context not related to the Internet of Money. They call the machine learning component the ``triage model'', tasked to process the rule-generated alerts: the generated score enables alert suppression or prioritization. 
The GuiltyWalker~\citep{oliveira2021guiltywalker} leverages random walks on a crypto-transaction graph to characterize distances to previous suspicious activity. 

Table~\ref{tab:tableComparison} shows a summary of the most influential research cited in this section. In this work, we aim to enhance the performance of classifier methods based on machine learning and graph analysis. To this end, we (i) adopt a novel scheme for transaction classification based on GAT; and (ii) resort to an updated implementation of GCN with respect to related works. As pointed out in the results section, this configuration improves state-of-the-art performance. 
Our methodology is backed up by an analysis of crypto-specific AML/CFT issues 
and anomaly detection approaches addressed in the next section.
In particular, we consider the set of transactions and their inherent characteristics, {\em i.e.} the fact that to spend cryptocurrencies, a user needs to have received them from previous transactions. These dependencies allow the creation of a graph whose structure can help identify illicit transactions. However, the need arises to 
identify the criteria that can inform a proper transactions classification -- {\em e.g.},
defining how it is possible to state that if a transaction is illicit, its neighbor transactions are also illicit, or if any graph-specific patterns represent suspicious activities. To confront these issues, it is essential to have a clear understanding of anomaly detection approaches in the RegTech field.

\section{Anomaly Detection Approaches}\label{sec:anomaly}

The process of anomaly/outlier detection involves processing data to detect behavior patterns that may indicate a change in system operations. The goal is to single out rare or suspicious events/items -- {\em i.e.}, those significantly different from the dataset~\citep{Kamisalic2021}. 
While collective anomaly detection methods target groups of data points that differ from most of the data, point anomaly detection also considers single data points~\citep{lixiang2022, sym14020328}. 
AML/CFT regulated entities, especially in the financial industry, deploy RegTech solutions to screen their operations and detect anomalous activities in an automated way. 
Their effort is based on the risk indicators 
provided by regulators usually in a rule-based format -- {\em i.e.}, templates of sequences of actions that suggest a suspicion in a way that is self-explainable and interpretable. Indeed, compliance decisions must be explainable and traceable for auditing. For this reason, the preliminary review of a flagged account relies on suspiciousness heuristics ({\em e.g.}, political exposure, geographic location, transaction type, users' behavior)~\citep{Weber2018}. This is the case of the mentioned FATF's indicators, developed from analyzing 100+ case studies from 2017 to 2020~\citep{FATF2020}.
Rule-based red flags can pertain to transaction patterns, such as `` incoming transactions from many unrelated wallets in relatively small amounts (accumulation of funds) with subsequent transfer to another wallet or full exchange for fiat currency'', or to anonymity, such as ``moving a VA that operates on a public, transparent blockchain, such as Bitcoin, to a centralized exchange and then immediately trading it for an AEC or privacy coin''~\citep{FATF2020}. 
{ In particular, indicators related to anonymity include cases of enhanced obfuscation ({\em e.g.}, AECs) and disintermediation ({\em e.g.}, unhosted wallets).}

In this context, a lot of time and resources are needed to investigate alerts generated by rule-matching processes and decide when to report a transaction as suspicious. 
An alert can be a true or a false positive, and arguably the simplicity of rule-based systems, despite guaranteeing interpretability, produces an estimate of around 95–98\% false positives~\citep{Eddin2021}.

Indeed, classifying entities and discovering patterns in massive time-series transaction datasets that are dynamic, high dimensional, combinatorially complex, non-linear, often fragmented, inaccurate, or inconsistent is a challenging task. Moreover, the difficulty of automating the synthesis of information from multi-modal data streams thrusts the task onto human analysts. 
This adds to a vicious circle of a compliance approach that stimulates over-reporting due to the cost asymmetry between false positives and false negatives and overburdens law enforcement agencies~\citep{Weber2018}.
Hence, the automation of an increasing array of processes has been suggested~\citep{electronics10151766}. 

Against this backdrop, in this section 
we explore the
anomaly detection methods that relate to our experiments. Hence, we focus on 
machine learning and graph analysis. We take an on-chain data analytic perspective, although we acknowledge the value of tools that target off-chain data, such as Natural Language Processing and sentiment analysis, that also leverage graph methods~\citep{Weber2018}. Indeed, while cryptocurrency transactional data is often analyzed through a combination of on-chain and off-chain techniques, thus including information not recorded on the blockchain or recorded on a different blockchain, in this work, we focus on on-chain data.

\subsection{Machine Learning}

Machine learning is a part of artificial intelligence that exploits data and algorithms to imitate human learning processes with gradual accuracy improvements. This helps us find solutions to problems in many fields, {\em e.g.}, vision, speech recognition, robotics~\citep{alpaydin2020introduction}.
In the most diverse contexts, it provides tools that can learn and improve automatically leveraging the vast amount of data available in our age~\citep{Kamisalic2021}.
In the compliance domain, advances in these algorithms show great promise, and their deployment in AML/CFT RegTech solutions can improve the efficiency of these applications~\citep{Weber2019}. For instance, they can mitigate the shortcomings of rules-based systems and infer patterns from historical data, increasing detection rates and limiting false positives \citep{lorenz2021machine}. 
In other cases, a more proactive approach is deployed to map and predict illicit transactions~\citep{Weber2019, Koshy2014}.

One of the main distinctions in machine learning is between unsupervised methods, where the model works on its own to discover patterns and information previously undetected, and supervised techniques, where labeled datasets are used to train algorithms. 
While applying both methods for anomaly detection is possible, most systems deploy unsupervised techniques due to a lack of relevant real-world labeled datasets. 
In the AML/CFT sphere, this scarcity mainly derives from difficulties in labeling real cases timely and comprehensively. Indeed, manual labels are costly in terms of time and effort, and the nature of the entities involved is complex and ever-evolving~\citep{lorenz2021machine}. 
Hence, 
analytic companies play a key role in labeling crypto transactions.
In order to address the overall lack of data, various strategies have been proposed~\citep{Eddin2021}: generate a fully synthetic dataset, simulate only unusual accounts within a real-world dataset, and localize rare events within a peer group. However, better validations of the systems were obtained using analyst feedback or real labeled data. 
Parallelly, the dataset shortage has driven the deployment of 
active learning ({\em i.e.}, few labels)~\citep{lorenz2021machine}.

\subsubsection{Supervised Baseline Techniques}
Supervised learning techniques are leveraged for their labeled training data. For instance, they are used to classify anomalies based on association rules to detect suspicious events~\citep{luo2014suspicious}. In the AML/CFT context, the label of each transaction could indicate whether it was identified as money laundering or not~\citep{lorenz2021machine}.
Recent RegTech solutions deploy widespread supervised learning methods to perform anomaly detection~\citep{Yin2019}:

\begin{itemize}
    \item \textbf{Decision Tree} - It is one of the base algorithms used in machine learning, with a name derived from a hierarchical model formed visually as a tree where nodes are decisions with specific criteria. The training data is subdivided into subsets following the tree branches. The node decision criteria are determined variables that can be defined as explanatory. The algorithm tries to apply the most significant feature to perform the best division among the training data. The best division can be measured by the information gain, mathematically derived from a decrease in entropy~\citep{alpaydin2020introduction}.
    
    \item \textbf{Random Forests} - It is an extension of Decision Trees in which an algorithm approaches the classification task by constructing a multitude of trees. Introduced by~\citep{breiman2001random}, it is an ensemble method applied to sample random subsets of the training data for each Decision Tree. It aims to improve the predictive accuracy of a classifier by combining multiple individual weak learners, {\em i.e.}, trees.
    
    \item \textbf{Boosting Algorithms} - They are another ensemble method that fits weak learners' sequences. A boosting algorithm tries to boost a Decision Tree by recursively selecting a subset of the training data. AdaBoost (Adaptive Boosting) assigns weights to the data samples based on the weak learners' ability to predict the individual training sample. Thus, the sample weights are individually computed for each iteration, and the successive learner is applied to the new data subset~\citep{Yin2019}. 
    
    \item \textbf{Logistic Regression} - It is a multiple regression suitable for binary classification, which assesses the relationship between the binary dependent variable (target) and a set of independent categorical or continuous variables (predictors)~\citep{hilbe2009logistic}. 
    It can be seen as measuring the probability of an event happening, where the probability consists of the ratio between the probability that an event will occur and the probability that it will not.
    
    \item \textbf{Support Vector Classification (SVC)} - Given a set of data for training, each labeled with the class to which it belongs among the two possible classes, a training algorithm for Support Vector Machines builds a model that assigns the new data to one of the two classes. This generates a non-probabilistic binary linear classifier. This model represents data as points in space, mapped in a way that a space separates data belonging to the two categories as ample as possible. New data is then mapped in the same space, and the prediction of the category to which they belong is made based on the side in which they fall~\citep{alpaydin2020introduction}. 
    
    \item \textbf{k-Nearest Neighbours (k-NN)} - It is a supervised learning algorithm used in pattern recognition for object classification based on the characteristics of the objects close to the considered one. The model represents data as points in space, {\em i.e.}, the feature space. Given a notion of distance between data objects, the input is the k nearest training data in the feature space. The underlying idea is that the more similar the instances, the more likely they belong to the same class~\citep{alpaydin2020introduction}. 
\end{itemize}


\subsubsection{Graph Analysis}
In recent years, a portion of machine learning research focused on real-world datasets that come in graphs or networks -- {\em e.g.}, social networks, knowledge graphs -- to generalize learning models to such structured datasets. Graph analytics is becoming increasingly important for AML/CFT, because money laundering involves flow relationships between entities that create graph structures. Some approaches for supervised learning work with graph-structured data based on a variant of neural networks which operate directly on graphs, {\em i.e.}, graph neural network~\citep{design_space,kipf2016semi}. Convolutional neural networks, for instance, offer an efficient architecture to extract significant statistical patterns in large-scale and high-dimensional datasets and can be generalized to graphs~\citep{kipf2016semi,defferrard2016convolutional}.
{
In this work, we use two specific graph-based neural networks, {\em i.e.}, Graph Convolutional Networks (GCN) and Graph Attention Networks (GAT). These techniques are described in the next section.
}

\section{Experimenting With Machine Learning}\label{sec:experiment}

As contextualized above, AML/CFT analytics benefit from deploying machine learning-based techniques for transaction classification. 
However, in new techniques, there is the need to balance interpretability and explainability with the reduction of false positives and over-reporting.
Accordingly, this section outlines the experimental setup of our study and the relevant results. After describing the dataset, we consider the evaluation method and the implementation of the anomaly detection approaches. Subsequently, we compare the results of our experiments, where state-of-the-art machine learning techniques and graph-based neural networks are employed in an AML/CFT context.

It is worth noting that, in developing this work, we heed several assumptions. Although we have already discussed these throughout the text, we provide the following summary. In our paper, (i) the term Internet of Money refers to the entire set of cryptocurrency ecosystems; (ii) we do not offer a comprehensive review of crypto forensic techniques; (iii) we focus on on-chain data; (iv) we perform inevitable simplifications when addressing explainability and interpretability of AI applications and relevant legal impacts.

\subsection{Methodology}
Our experimentation is grounded on a seminal work by~\citep{Weber2019}. 
{
Most of the techniques deployed in the study correspond to the standard supervised models mentioned above -- {\em i.e.}, Decision Trees, Logistic Regression, k-NN, SVC, AdaBoost, Random Forests --, used as benchmark methods for classification.
However, the two graph-based models GCN and GAT deserve close attention in this context. This is for three main reasons:
(i) these types of neural networks take into account the graph nature of our dataset; (ii) as the evaluation shows, our application of GCN outperforms benchmark approaches and improves the state of the art; (iii) to the best of our knowledge, this is the first attempt to deploy the GAT model in the AML/CFT context.}

\subsubsection{Transaction Graph Analysis}
Graphs represent a typical mathematical tool to model interactions among different entities: humans, elements of a biological system, computing nodes in a distributed system, and others
~\citep{doubleblind}. 
In a blockchain, transactions are linked by nature since money spent in a transaction originates from previous transfers 
~\citep{doubleblind}. 
This allows the creation of a graph of transactions that can help the classification process. In fact, given a transaction $t$, it is possible to collect all the connected transactions and recursively search for other ones up to a certain depth level. Given such a connected graph centered at $t$, an inspection of the neighboring transactions and their classified value can aid the classification of $t$.
Each node of the graph (transaction) has thus a set of neighbors that will influence its classification. 
Moreover, each node has a set of features associated with the corresponding transaction (see below for the details of the dataset).

\begin{figure}[htp]
    \centering
    \includegraphics[width=.6\textwidth]{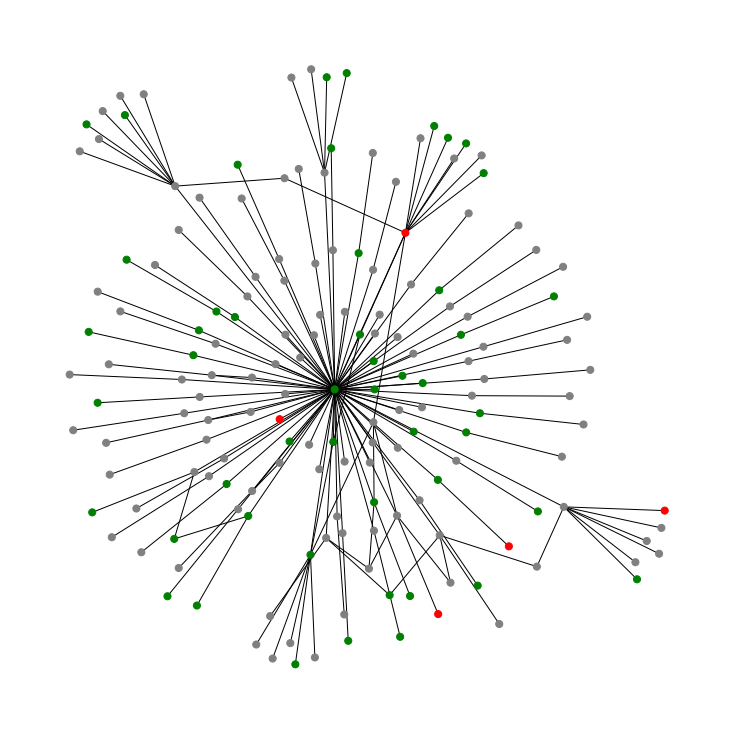}
    \includegraphics[width=.6\textwidth]{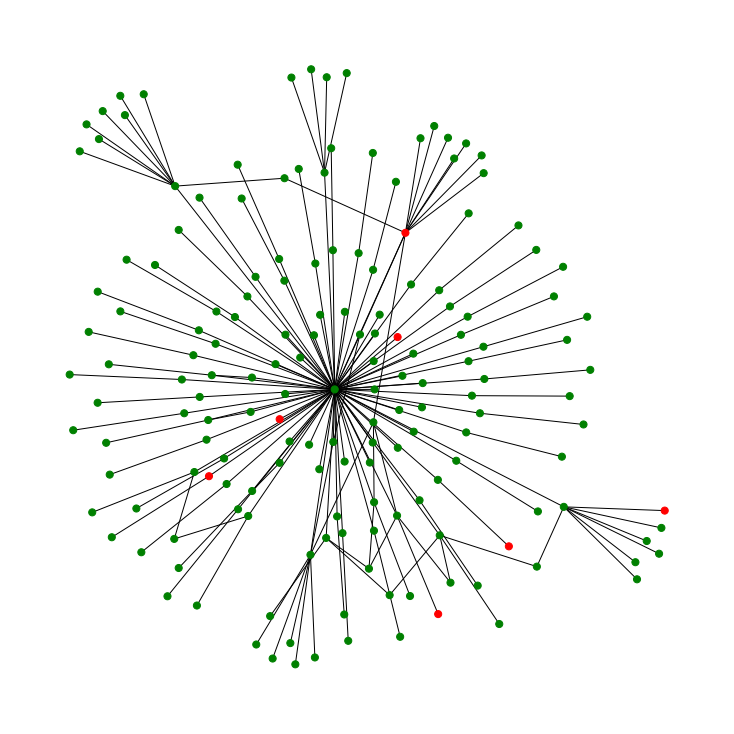}
    \caption{Connected graph of a considered transaction before and after classification.}
    \label{fig:tracker_comparison}
\end{figure}

An example of this procedure is displayed in Figure \ref{fig:tracker_comparison}, where a connected component -- {\em i.e.}, a subgraph in which each pair of nodes is connected via a path -- is obtained from an initial transaction (Figure \ref{fig:tracker_comparison}, top). In the Figure, {\em red} nodes represent transactions labeled as illicit in the starting dataset, the {\em green} ones licit transactions and the {\em grey} ones are 
still unknown/unlabeled. 
To show the output of a machine learning classification problem, the bottom part of Figure \ref{fig:tracker_comparison} shows the output of the process employing a specific classification algorithm, which in this case is Random Forest. 
In essence, the idea is that knowing the labels of certain transactions aids the classification of the remaining (unknown/unlabeled) ones. 
Hence, learning methods could pinpoint illicit transactions
based on the graph topology and the features of the transactions.

\subsection{Dataset}

In our work, we experimented with the publicly available Elliptic transactions dataset provided in the context of~\citep{Weber2019}. 
{ For details on the dataset, the reader can refer to the latter and to the description provided on Kaggle together with the dataset\footnote{Elliptic dataset: https://www.kaggle.com/datasets/ellipticco/elliptic-data-set}. This dataset contains real Bitcoin transactions represented as a directed graph network, where transactions are nodes, and the directed edges between these transactions represent fund flows from the source address to the destination address. 
The dataset contains 203,769 transaction nodes connected by 234,355 edges. For each transaction, 167 features are available, of which the first 94 relate to the transaction itself and thus directly extracted from the blockchain -- {\em e.g.}, the number of inputs of a transaction or the number of outputs -- while
the other 73 features relate to the graph network itself and are extracted from the neighboring transactions of a node. 
The features do not have any 
associated descriptions -- indeed, the authors in \citep{Weber2019} claim that they cannot describe these features due to intellectual property issues.}
Tests were carried out with transaction features (tx) and transaction features plus aggregated features (tx + agg). Such aggregated features are obtained by aggregating transaction information one-hop backward/forward from the center transaction node. This means obtaining the features of the nodes that share an edge with that transaction node. 

Each transaction in the dataset is labeled as illicit, licit, or unknown: 4,545 are labeled as illicit, 42,019 are labeled as licit, and the remaining 157205 are unknown.
The transactions also contain temporal data. In particular, this is grouped into 49 distinct time steps, evenly spaced the interval of two weeks. Each time step contains a connected graph that includes all the transactions verified on the blockchain in the span of 3 hours~\citep{Weber2019}.


\begin{table}[htp]
    \centering
    \caption{GCN model architecture. Total parameters = 18,774, Trainable parameters = 17,756, Non-Trainable = 1,018.}
    \begin{tabular}{|c|c|c|}
        \hline
        \textbf{Layer (type)} &  \textbf{Output Shape} & \textbf{Num. parameters} \\
        \hline
        \hline
        preprocess (Sequential) & (46564, 32) & 4564 \\ 
        \hline
        convolution 1 (GraphConvLayer) & multiple & 5888 \\ 
        \hline
        convolution 2 (GraphConvLayer) & multiple & 5888 \\ 
        \hline
        postprocess (Sequental) & (46564, 32) & 2368 \\ 
        \hline
        logits (Dense) & multiple & 66 \\ 
        \hline
    \end{tabular}
    \label{tab:gcn_model}
\end{table}

The dataset was pre-processed as follows: (i) the features were merged with the classes; (ii) 
class values were renamed to integer values; (iii) transaction identifiers were swapped for a sorted index; (iv) only the part of the dataset labeled licit or illicit was selected; (v) all the edges between unknown transactions were removed. 
After the pre-processing, our cleaned dataset encompassed 46,564 transactions and 36,624 edges.

\subsection{Graph Convolutional Network Model Architecture}

{
The objective of a GCN model is to learn a function of signals/features on a data set structured as a graph. The model takes as input (i) a graph with nodes and edges between nodes and (ii) a feature description for each node. The key idea is that each node receives and aggregates features from its neighbors to represent and compute its local state. The GCN then usually produces an output feature matrix at the node level~\citep{kipf2016semi}.}
The GCN model is used for transaction classification because it is a deep neural network that allows capturing the relation among the nodes and their neighborhoods. In other words, it creates a node embedding in a latent vector space that captures the characteristics of the node neighborhood in the graph. This information comes in the form of a look-up table mapping nodes to a vector of numbers.
GCNs have been developed using the Keras framework, following the recommendations introduced in \citep{design_space}. 

The general structure of our graph convolution layer is made of three steps. First, the input node representations are processed using a Feed Forward Network to produce a message. Second, the messages of the neighbors of each node are aggregated using a permutation invariant pooling unsorted segment sum operation. Third, the node representations and aggregated messages are combined and processed to produce the new state of the node representations (node embeddings) via concatenation and Feed Forward Network processing.

Our network architecture consists of a sequential workflow of the model that we display in Table \ref{tab:gcn_model} and summarized as follows:
\begin{enumerate}
    \item Apply pre-processing using Feed Forward Network to the node features to generate initial node representations;
    \item Apply two graph convolutional layers, with skip connections, to the node representation to produce node embeddings;
    \item Apply post-processing using Feed Forward Network to the node embeddings to generate the final node embeddings;
    \item Feed the node embeddings in a Softmax layer to predict the node class.
\end{enumerate}


\begin{table}[htp]
    \centering
    \caption{GAT model architecture. Total parameters = 59,952, Trainable parameters = 59,952, Non-Trainable = 0.}
    \begin{tabular}{|c|c|c|}
        \hline
        \textbf{Layer (type)} &  \textbf{Output Shape} & \textbf{Num. parameters} \\
        \hline
        \hline
        dense 9 (Dense) & multiple & 10340 \\ 
        \hline
        dropout 6 (Dropout) & multiple & 0 \\ 
        \hline
        graph attention (MultiHeadGraphAttention) & multiple & 12320 \\ 
        \hline
        dense 10 (Dense) & multiple & 36630 \\ 
        \hline
        dropout 7 (Dropout) & multiple & 0 \\ 
        \hline
        dense 11 (Dense) & multiple & 662 \\ 
        \hline
    \end{tabular}
    \label{tab:gat_model_architecture}
\end{table}

\subsection{Graph Attention Network Model Architecture}

{
While the GCN model averages the node states from source nodes to the target node, the GAT model gives different importance to each node's edge by using an attention mechanism to aggregate information from neighboring nodes \citep{Velickovic2017-bc}. In other words, instead of simply averaging/summing node states from source nodes to the target node, as we do in the GCN model, GAT, on the other hand, first applies normalized attention scores to each source node state and then sums~\citep{Velickovic2017-bc}.
}

Our model is built using the Keras framework that, through a graph attention layer that computes pairwise attention scores, aggregates and applies the scores to the node's neighbors. A multi-head attention layer concatenates multiple graph attention layer outputs. 
Our design choice is to use a single attention layer with multiple heads, enabling the network to jointly attend multiple positions~\citep{multi_head}. The multi-head layer is then inserted into a general model that implements dense pre-processing/post-processing layers with dropout regularization, as shown in Table \ref{tab:gat_model_architecture}. 
The training proved to be subjected to overfitting, and heavy regularization was necessary, which was achieved by dropout layers and using RMSprop optimizer with momentum \citep{Philipp2017-yh}.


\begin{table}[htp]
    \centering
    \caption{Table showing the results for { \textbf{illicit} transaction classification with} the F1-score, Micro Average F1-score, Precision and Recall metrics for all models.}\label{tab1}
    \begin{tabular}{|c|c|c|c|c|}
    \hline
    \textbf{Model} &  \textbf{Precision} &  \textbf{Recall} &  \textbf{F1 Score} &  \textbf{M.A. F1} \\
    \hline
    \hline
    Random Forest Classifier (tx) &      0.909 &   0.648 &     0.757 &         0.974 \\
    \hline
    Random Forest Classifier (tx + agg) &      0.981 &   0.651 &     0.782 &         \textbf{0.977} \\
    \hline
    Logistic Regression (tx) &      0.515 &   0.646 &     0.573 &         0.939 \\
    \hline
    Logistic Regression (tx + agg) &      0.456 &   0.630 &     0.529 &         0.929 \\
    \hline
    MLP (tx) &      0.897 &   0.593 &     0.714 &         0.970 \\
    \hline
    MLP (tx + agg) &      0.817 &   0.623 &     0.707 &         0.968 \\
    \hline
    k-NN Classifier (tx) &      0.762 &   0.629 &     0.689 &         0.964 \\
    \hline
    k-NN Classifier (tx + agg) &      0.730 &   0.576 &     0.644 &         0.960 \\
    \hline
    SVC (tx) &      0.842 &   0.604 &     0.703 &         0.968 \\
    \hline
    SVC (tx + agg) &      0.862 &   0.588 &     0.699 &         0.968 \\
    \hline
    Decision Tree Classifier (tx) &      \textbf{0.986} &   0.573 &     0.725 &         0.973 \\
    \hline
    Decision Tree Classifier (tx + agg) &      \textbf{0.986} &   0.573 &     0.725 &         0.973 \\
    \hline
    AdaBoost Classifier (tx) &      0.793 &   0.615 &     0.693 &         0.966 \\
    \hline
    AdaBoost Classifier (tx + agg) &      0.945 &   0.567 &     0.708 &         0.971 \\
    \hline
    GCN (tx) &      0.906 &   \textbf{0.790} &     \textbf{0.844} &         0.973 \\
    \hline
    GAT (tx) &      0.897 &   0.605 &     0.723 &         0.971 \\
    \hline
    \end{tabular}
\end{table}

\begin{figure}[htp]
    \centering
    \includegraphics[width=.65\textwidth]{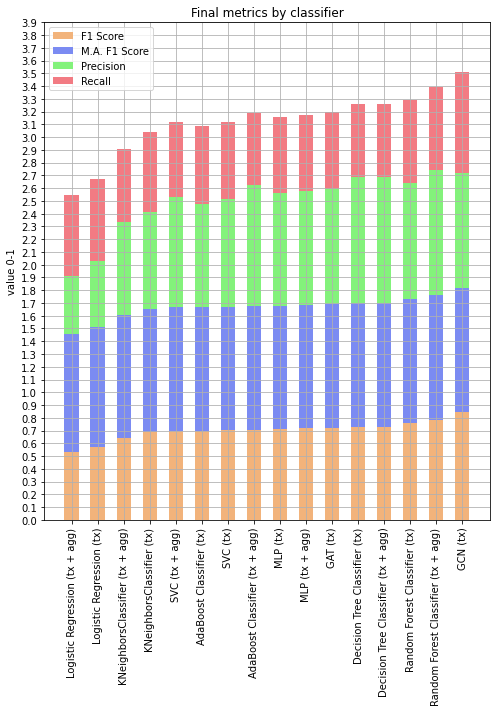}
    \caption{Barplot aggregating F1-score, Micro Average F1-score, Precision and Recall for all the approaches experimented.}
    \label{fig:gcn_table}
\end{figure}

\subsection{Results}

{
For the discussion of the results, we firstly consider the illicit class as, due to the nature of the dataset (less labeled illicit transactions) and of the problem, its classification is more complex. 
To compare the results, we use the F1-score,}
a metric obtained from Precision and Recall. 
{
These metrics are usually defined for a binary classifier (as in this case) where some special instances need to be identified, {\em e.g.}, positive cases to a particular test.
Precision is the number of true positive (TP) predictions, {\em i.e.}, how many of the positive predictions made are correct over the sum of TP and false positives (FP). In other words, precision says how many of the identified illicit transactions were illicit.
$$ \text{Precision} = \frac{\text{TP}}{\text{TP+FP}}.$$
Recall measures how many positive cases the classifier correctly predicted over all the positive cases in the data, {\em i.e.}, TP and false negatives (FN). In our context, for instance, it allows us to understand how many illicit transactions the classifier identified over the real considered set of illicit transactions.
$$ \text{Recall} = \frac{\text{TP}}{\text{TP+FN}}.$$
}
The F1-score represents the harmonic mean of Recall and Precision and is thus calculated as:

$$ \text{F1} = 2 \times \frac{\text{Precision} \times \text{Recall}}{\text{Precision} + \text{Recall}} $$

We also use the Micro Average F1-score for evaluating the methods. It measures the F1-score of the aggregated contributions of all classes.

The final performance results are reported in Figure \ref{fig:gcn_table} and Table \ref{tab1}.
It is possible to observe how GCN outperforms other approaches. 
In particular, the GCN approach provides the best results in terms of recall, {\em i.e.}, $0.790$, and F1-score, {\em i.e.}, $0.844$. In terms of precision, it slightly deviates from the Decision Tree ($0.986$) and Random Forest ($0.981$) approaches but still provides better performances than all the rests, {\em i.e.}, $0.906$. 
For what concerns the Micro Average F1-score, all approaches fit in the range of $0.960$ to $0.977$. These results are in contrast with the results in~\citep{Weber2019}, where Random Forests provided the best performance. 
The 
{ cause of} 
such improvement might be due to the different architectures we exploited to build the neural network. 

Furthermore, the comparison with the other graph-based approach, {\em i.e.}, GAT, also sees the GCN outperforming. GAT performs better than a simple dense network but cannot reach the results of GCN and Random Forest classifiers. The motivation is probably due to the naïve structure of the neural network, and its optimization is currently under investigation.


{

\begin{table}[htp]
    {
    \centering
    \caption{Table showing the results for { \textbf{licit} transaction classification with} the F1-score, Micro Average F1-score, Precision and Recall metrics for all models.}\label{tab2}
    \begin{tabular}{|c|c|c|c|c|}
    \hline
    \textbf{Model} &  \textbf{Precision} &  \textbf{Recall} &  \textbf{F1 Score} &  \textbf{M.A. F1} \\
    \hline
    \hline
    Random Forest Classifier (tx) &      0.977 &   0.995 &     0.986 &         0.973 \\     
    \hline
    Random Forest Classifier (tx + agg) &      0.977 &   \textbf{0.999} &     \textbf{0.988} &         \textbf{0.978} \\     
    \hline
    Logistic Regression (tx) &      0.976 &   0.959 &     0.967 &         0.939 \\     
    \hline
    Logistic Regression (tx + agg) &      0.975 &   0.949 &     0.962 &         0.929 \\     
    \hline
    MLP (tx) &      0.973 &   0.995 &     0.984 &         0.970 \\     
    \hline
    MLP (tx + agg) &      0.974 &   0.994 &     0.984 &         0.970 \\     
    \hline
    k-NN (tx) &      \textbf{0.978} &   0.967 &     0.972 &         0.949 \\     
    \hline
    k-NN (tx + agg) &      0.975 &   0.965 &     0.970 &         0.944 \\     
    \hline
    SVC (tx) &      0.974 &   0.992 &     0.983 &         0.968 \\     
    \hline
    SVC (tx + agg) &      0.973 &   0.994 &     0.983 &         0.968 \\     
    \hline
    Decision Tree Classifier (tx) &      0.972 &   \textbf{0.999} &     0.986 &         0.973 \\     
    \hline
    Decision Tree Classifier (tx + agg) &      0.972 &   \textbf{0.999} &     0.986 &         0.973 \\     
    \hline
    AdaBoost Classifier (tx) &      0.975 &   0.989 &     0.982 &         0.966 \\     
    \hline
    AdaBoost Classifier (tx + agg) &      0.972 &   0.998 &     0.985 &         0.971 \\      
    \hline
    GCN (tx) &      0.975 &   0.994 &     0.984 &         0.971 \\  
    \hline
    GAT (tx) &      0.973 &   0.992 &     0.982 &         0.967 \\   
    \hline
    \end{tabular}
    }
\end{table}

So far, we have focused on the classification of illicit transactions. Since the dataset is unbalanced -- {\em i.e}., it contains more licit than illicit transactions in a ratio of more or less 1 to 10 -- the problem becomes relatively trivial. For the sake of transparency, however, the final performance results related to licit transactions are reported in Table \ref{tab2}.
All the models perform very well and are very similar to each other. In this case, graph-based approaches do not perform better than Random Forest classifiers, but the difference is not outstanding.
}

\section{Discussion}\label{sec:discussion}

When interpreted through the lens of the AML/CFT remarks outlined in the previous sections, our findings inspire multi-layered considerations. 
{ Accordingly, in this section our reasoning is threefold. First, we discuss the results of our experiments vis-à-vis the approaches used as benchmarks. Secondly, we broaden the perspective of the analysis to consider not only the impacts of crypto-related RegTech methodologies on the evolution of the Internet of Money, but also the interplay between the latter and the prospective role of forensics. Finally, we pinpoint a few associated challenges.}

{ From the first perspective,} to the best of our knowledge the experiment described in this paper is the first attempt to implement GAT models to detect anomalies in Bitcoin transactions for AML/CFT purposes.
The final results are on par with the state of the art of GCN networks, with GAT marginally worse than GCN. 
This could be explained by the ``simpler'' implementation of GAT and the possibility that the dataset responds better to non-spectral methods.
Nonetheless, we argue that the novelty of this application could be helpful for general research on GAT anomaly detection techniques. 
In addition, the results show that the GCN neural network typology is a promising solution for AML/CFT, as it performs better than other approaches.

In this context, it is essential to consider that GCN and GAT classifiers only have access to transaction features, which means that all information about aggregated nodes comes from the graph structure itself. Since the performance of GCN is in line with Random Forest (with aggregated features), we can claim that our graph networks can obtain the same amount of information as the creator of the dataset~\citep{Weber2019}.
However, choosing one method over another carries additional implications that must be carefully weighed. 
For example, the performance of Random Forests falls slightly behind GCN's, but there is no sacrifice in explainability because the detectors are derived from Random Forests' rules~\citep{Eddin2021}. 
Given the size and dynamism of real-world information, explainability of the results is challenging to provide, both in this context and in the broader AI field. 
Even in our specific narrow instance -- 
{\em i.e.}, transaction graphs that model illicit activity over time -- it is challenging to apply efficient methods whose results can be understood by humans. 
Although this appears to be a crucial aspect, the literature still lacks some research on the application of explainable AI techniques for AML/CFT anomaly detection~\citep{kute2021deep}.

{ From the second perspective, the choice of the forensic approach(es) to deploy must be made taking into consideration the evolution of the Internet of Money, with specific regard to peer-to-peer transfers and DeFi protocols.
Indeed, while its developments warrant the application of increasingly sophisticated yet explainable compliance and investigation techniques, we see how} the implementation of the crypto travel rule has already prompted the industry to denounce the lack of global standards and technical solutions to underpin effective and affordable compliance. 
{ It follows that, while the great quantity and complexity of transaction data to be processed 
suggests that machine learning will continue to be a part of the solution -- with marginal performance differences possibly bearing significant weight when various approaches are combined --, it is crucial to back the relevant research with a constructive dialogue between the stakeholders involved.}
In this context, we point out to
the increase in the laundering-related use DeFi protocols of 1.964\% between 2020 and 2021. In 2021, centralized exchanges received 47\% of funds originating from illicit addresses and DeFi protocols 17\%, vis-à-vis 2\% in 2020. Likewise, in 2021 funds derived from cryptocurrency thefts were increasingly sent to DeFi platforms (51\%) or risky services (25\%), while only 15\% went to centralized exchanges, possibly due to AML/CFT~\citep{ChainalysisTeam2022}. 
{ In 2022, almost half of illicit crypto funds passed through a set of intermediary services primarily populated by mixers, illicit services and DeFi protocols.
However, 67\% of illicit funds received by exchanges went to only five centralized exchanges, in comparison to 56.7\% of 2021~\citep{ChainalysisTeam2023}.}

{ In the near future,
regulated entities, law enforcement and supervisors will increasingly need to monitor and analyze crypto transactions to which  multi-layered obfuscation techniques have been applied. In addition, given 
the rise in the use of unhosted wallets and decentralized platforms, they will frequently 
operate 
without the assistance of centralized counterparty entities.}
For these reasons, we wish to highlight the value of not only researching innovative machine learning-based forensic applications, but also to adopt an interdisciplinary approach to devise compliance tools that adequately consider the way regulatory regimes are conceived and enforced. { For instance, we point to the importance of 
reconciling the duties placed on regulated entities, the available and prospective intelligence tools, and an AML/CFT regime that is so far inherently and explicitly intermediary-based, with compliance efforts guided by rule-based risk indicators. It is for this reason that} 
our work contextualizes forensic methods into the specifics of risk indicators. 
{ Building on these arguments,}
we emphasize that AML/CFT hurdles cannot be solved by simply resorting to a sophisticated transaction classification scheme. 
On the contrary, this process needs to be nested into a broader framework to be effective. 

Indeed,
{ our analysis of machine learning methods was anchored to the mitigation of the drawbacks of current rule-based systems in terms of} false positives and over-reporting.
Relatedly, we find that the value of experimenting with machine learning algorithms for RegTech purposes appears dependent mainly on the relationship between the given approach and the regulatory environment within which it is deployed. In other words, the efficiency of a specific algorithm can be assessed {\em per se}, but its effectiveness in an AML/CFT context heavily depends on the extent to which the structure of the model correctly mirrors the regulatory framework -- {\em e.g.}, it generates alerts that are deemed relevant by regulators and mitigates the current trends of over-reporting. 

{ From the third perspective, 
the supervised classification analysis we conducted could be in theory applied to other types of blockchain and cryptocurrencies, being the analysis constrained on the high-level perspective of the cryptocurrency transactions' graph. However, there is the need for a labeled transaction dataset to build such a transaction graph. And the lack of open data further complicates the task.
Indeed, we point out a few challenges identified during our investigation, related to the openness and availability of the datasets being discussed and the explainability of the results. 
We find it is overarching to confront these open issues
and devise appropriate solutions or mitigating measures. On the one hand, our analysis suggests that it is difficult to address efficiency evaluations of machine learning-based AML/CFT tools for anomaly detection and transaction classification, since this feature appears to be increasing to the detriment of interpretability and explainability.
On the other hand, it is evident from our studies that the labelled transaction datasets on which supervised learning algorithms are trained are largely proprietary. 
This does not only impact the development of new methods, but possibly also the transparency of the activity of supervisory bodies. That is, if the activity of the latter, just as the compliance effort of regulated entities, can be based only on the intelligence findings of solutions deploying proprietary algorithms.
The interplay between the lack of explainability and the proprietary nature of the datasets suggests worrisome scenarios that call for further research. 
Hence, it is crucial to foster public-private synergies that can consider the AML/CFT context 
from a socio-technical, operational and regulatory viewpoint.}


\section{Conclusions}\label{sec:concl}
 
{ Elaborating on the enthusiasm for the financial application of blockchain and DLTs that surged in the wake of Bitcoin's launch, today the Internet of Money comprises a diverse set of socio-technical systems under constant evolution -- {\em e.g.}, recently, DeFi schemes. 
Over the years, forensic techniques have been deployed to connect crypto addresses/transactions to real-world identities. This responds to the regulatory quest to ensure accountability through identification, a concept that sits at the core of AML/CFT compliance. Meanwhile, institutions and authorities drafted anomaly indicators to help with the identification of suspicious transfers in compliance with the risk-based approach.
In this context, law enforcement agencies and supervisors, often supported by the private sector, began to apply forensic methods to track relevant transfers, as well as regulated entities started benefiting from innovative RegTech solutions that partially automate the detection of anomalous activities.}

In this paper we focused on 
these techniques from an on-chain data analytic perspective, with a specific focus on approaches based on machine learning and graph analysis. 
The use of these algorithms in AML/CFT RegTech solutions shows great promise to improve the efficiency { of the latter and mitigate the significant drawbacks of current rule-based methodologies.}
To the best of our knowledge, what we described in this work is the first experiment with GAT models for AML/CFT anomaly detection in Bitcoin.
{ The application of this type of neural network falls in line with the recent focus on deploying machine learning techniques that leverage the inherent structure of many real-world datasets that come in the form of graphs or networks.}
GCN and GAT models are informed by the idea of creating generalized learning models for these structured datasets, and indeed the one we analyzed consists of 
(real) Bitcoin transactions represented as a directed graph network. 

{ To conclude, we provide three levels of considerations.}
From an operational standpoint, our results show that the mentioned graph-based methods perform better than the baseline approaches -- {\em e.g.}, GCN performs better than Random Forests, with GAT being marginally worse than GCN. This
{ encourages further experimentations with the use of GCN neural networks for AML/CFT purposes, while the novelty of our approach could spur}
further research into GAT-based anomaly detection techniques.
From a related methodological perspective, 
{ we argue that a constant experimentation with various forensic methods, possibly leveraging the value added by transaction graphs, is crucial to 
reap the full benefits of analytics in an ever-evolving context of application such as the Internet of Money.
These explorations, however,
must be backed by serious efforts to 
foster 
constructive public-private dialogue regarding the openness and the availability of labelled transaction datasets.}

{ From a final conceptual viewpoint,
we emphasize that a holistic interpretation of the interplay between AML/CFT measures and the Internet of Money -- {\em i.e.}, one that heeds in a comprehensive fashion socio-technical, operational and regulatory dynamics when defining the object of the analysis -- is crucial to devise effective and possibly efficient RegTech solutions. 
Indeed, the efficiency of a specific algorithm may not guarantee its effectiveness in an AML/CFT context, which depends on the extent to which the model responds to regulatory needs and generates relevant alerts. This relevance is influenced by regulatory, compliance and supervisory needs, as affected by the evolution of the features of the Internet of Money.
This holistic approach is especially valuable when it comes to transaction classification and anomaly detection, where a main challenge is the need to balance interpretability and explainability with the goal to reduce the share of false positives and over-reporting.}

\section*{Acknowledgements}
The contribution of Nadia Pocher and Mirko Zichichi received funding from the EU H2020 research and innovation programme under the MSCA ITN European Joint Doctorate grant agreement No 814177 Law Science and Technology Joint Doctorate - Rights of the Internet of Everything.

\bibliography{biblio}

\end{document}